\documentclass[apj]{emulateapj}

\slugcomment{ApJL, in press}

\shorttitle{Stellar disk truncations at high--z}

\shortauthors{Trujillo \& Pohlen}

\begin{document}

\title{Stellar disk truncations at high--z: probing inside--out galaxy formation}

\author{{Ignacio Trujillo\altaffilmark{1}} and Michael {Pohlen}\altaffilmark{2}} 

\altaffiltext{1}{Max--Planck--Institut f\"ur Astronomie, K\"onigstuhl 17, D--69117 Heidelberg,
Germany}
\altaffiltext{2}{Kapteyn Astronomical Institute, University of Groningen, NL-9700 AV
Groningen, The Netherlands}

\begin{abstract}

We have conducted a systematic search for stellar disk truncations in disk--like
galaxies at intermediate redshift (z$<$1.1) using the Hubble Ultra Deep Field (UDF)
data. We use the position of the truncation as a direct estimator of the size of
the stellar disk. After accounting for the surface brightness evolution of the
galaxies, our results suggest that the radial position  of the truncations has
increased with cosmic time by $\sim$1--3 kpc in the last $\sim$8 Gyr. This result
indicates a small to moderate ($\sim$25\%) inside--out growth of the disk galaxies
since z$\sim$1.

\end{abstract}

\section{Introduction}

Understanding the formation and evolution of galactic disks is an important
goal of current cosmology. The surface brightness of the disks of present--day
galaxies are well described by an exponential law (Freeman 1970), with a
certain scalelength, taken as the characteristic size of the disk. However,
since van der Kruit (1979) it is known that some stellar disks  are truncated
in the outer parts (for a recent review see Pohlen et al. 2004). Stellar disk
truncations have been explained as a consequence of the inhibition of
widespread star formation below a critical gas surface density (Kennicutt 1989;
Martin \& Kennicutt 2001). Alternatively, van der Kruit (1987), has proposed
that the truncation radius corresponds to the material with the highest
specific angular momentum in the protogalaxy. Recent models (Elmegreen \&
Parravano 1994; Schaye 2004) emphasize the transition to the cold interstellar
medium phase as responsible for the onset of local gravitational instability,
triggering star formation.

To date it is not systematically explored whether the above description of
present--day galactic disks is valid at high--z, and consequently, it is not clear
whether the truncation radius of high--z disks evolves with cosmic time. However,
very recently, P\'erez (2004) has shown that such kind of analysis is feasible at
intermediate redshift (z$\lesssim$1). Addressing the question of how the radial
truncation evolves with z  is strongly linked to our understanding of how the
galactic disks grow and how the star formation is taking place.
In this paper we propose the use of the truncation of stellar disks as a direct
estimator of their sizes. The evolution of the position of the
truncation, consequently, will clarify whether the galaxies are growing inside-out
with the star formation propagating radially outward with time.

The above issue can be addressed observationally with a structural analysis of galaxy
samples both in the near and in the high--z universe. Such studies require to follow
the galaxy surface brightness distributions down to very faint magnitudes
($\sim$27--28 V--band mag/arcsec$^2$) with enough signal to noise to assure a
reliable measurement of the disk truncation. The depth of the images plays a critical
role in the high--z truncation detections because of the cosmological surface
brightness dimming. In addition, it is necessary to work with high resolution images
to avoid the seeing effects on the shape of the surface brightness distribution. This
means to use the deepest  observations taken with the HST telescope. For that reason
we have decided to analyse the disk--like galaxies in the UDF. In order to maintain
our analysis in the optical restframe, the combination of k--correction and the
cosmological dimming restricts the use of the optical HST imaging for this project up
to z$\sim$1. Throughout, we will assume a flat $\Lambda$--dominated cosmology
($\Omega_M$=0.3, $\Omega_\Lambda$=0.7 and H$_0$=70 km s$^{-1}$ Mpc $^{-1}$).

\section{Data, Sample Selection and Surface Brightness Profiles}

Our galaxies have been selected in the Hubble Ultra Deep Field (Beckwith et al.
2005). This survey is a 400-orbit  program to image a single field with
the Wide Field Camera (WFC) of the Advanced Camera for Surveys (ACS) in four
filters: F435W (B), F606W (V), F775W (i), and F850LP (z). We have used the
public available V, i and z-band  mosaics with a pixel scale of 0.03$''$/pixel.
The  FWHM is estimated to be 0.09 seconds of arc.

To make the selection of our sample  we have taken advantage of the fact that the
UDF field is within the   Galaxy Evolution from Morphology and SEDs (GEMS; Rix et
al. 2004) imaging survey with the ACS pointing to the Chandra Deep Field South
(CDF--S).  Focusing on the redshift range 0.1$\leq$z$\leq$1.1, GEMS provides
morphologies and structural parameters for nearly 10000 galaxies (Barden et al.
2005; McIntosh et al. 2005). For these galaxies photometric redshift estimates,
luminosities, and SEDs exist from COMBO--17 (Classifying Objects by Medium--Band
Observations in 17 Filters; Wolf et al. 2001, 2003). The COMBO--17 team has made
this information publically available  through a catalog with precise redshift
estimates ($\delta$z/(1+z) $\sim$ 0.02) for approximately 9000 galaxies down to
m$_R$$<$24 (Wolf et al. 2004). Rest--frame absolute magnitudes and colors, accurate
to $\sim$0.1 mag, are also available for these galaxies.

Barden et al. (2005) have conducted the morphological analysis of the late--type
galaxies in the GEMS field by fitting S\'ersic (1968) r$^{1/n}$ profiles to the
surface brightness distributions. Ranvidranath et al. (2004) have shown that using
the S\'ersic index $n$  as a  criteria,  it is feasible to disentagle between late--
and early-- type galaxies at intermediate redshifts. Late--types (Sab--Sdm) are
defined through $n$$<$2--2.5. Moreover, the morphological analysis conducted  by
Barden et al. provides  the information about the inclination of the galaxies.
This is particularly important since we want to study the truncations of the
stellar disks in objects with low inclination. The edge--on view facilitates the
discovery of truncations but introduces severe problems caused by the effects of dust
and line--of--sight integration that we want to avoid (Pohlen et al. 2002).

Our sample is selected as follows, we have taken all the galaxies in the COMBO-17
CDF--S catalog with R$<$24 mag that are in common with the UDF object detection
catalog (Beckwith et al. 2005). This leaves a total of 166 objects. From these, 133
have photometric redshift estimation (i.e. a $\sim$80\% completeness). To maintain
our analysis in the optical restframe we select only the 118 galaxies with z$<$1.1.
The mean redshift of these galaxies is $\sim$0.68. At that redshift, the faintest  
absolute B--band restframe magnitude able to be analysed is M$_B$$\sim$-18.6 mag (AB
system). We take this value as a compromise between maximizing the number of objects
in our final sample and at the same time assuring as much as possible homogeneity
(equal luminous objects) through the full redshift range. Applying this magnitude cut
we get 63 galaxies. It is important to note however that, strictly, only those
objects brighter than M$_B$$<$-20 are observable homogeneously up to z=1.1. Finally,
we select from the 63 galaxies only those objects that according to the GEMS
morphological analysis have n$<$2.5 (i.e. those which have disk--like surface
brightness profiles) and e$\leq$0.5 (i.e. i$\leq$60$^\circ$). Our final sample
contains 36 galaxies. These galaxies are presented in Table \ref{table1}. 

To analyse the surface brightness profiles of our galaxies in a similar
rest--frame band along the explored redshift range (0.1$<$z$<$1.1), we have
extracted the profiles in the following  bands:  V--band for galaxies with
0.1$<$z$<$0.5, i--band for 0.5$<$z$<$0.8 and z--band for 0.8$<$z$<$1.1. This
allows us to explore the surface brightness distribution in a  wavelength close
to the B--band restframe. In addition, to probe whether the position of the
truncation depends on the observed wavelength we have analysed all galaxies in
the reddest band available: the z--band.

\begin{figure}
\vspace*{-1.3cm}
\includegraphics[width=10cm]{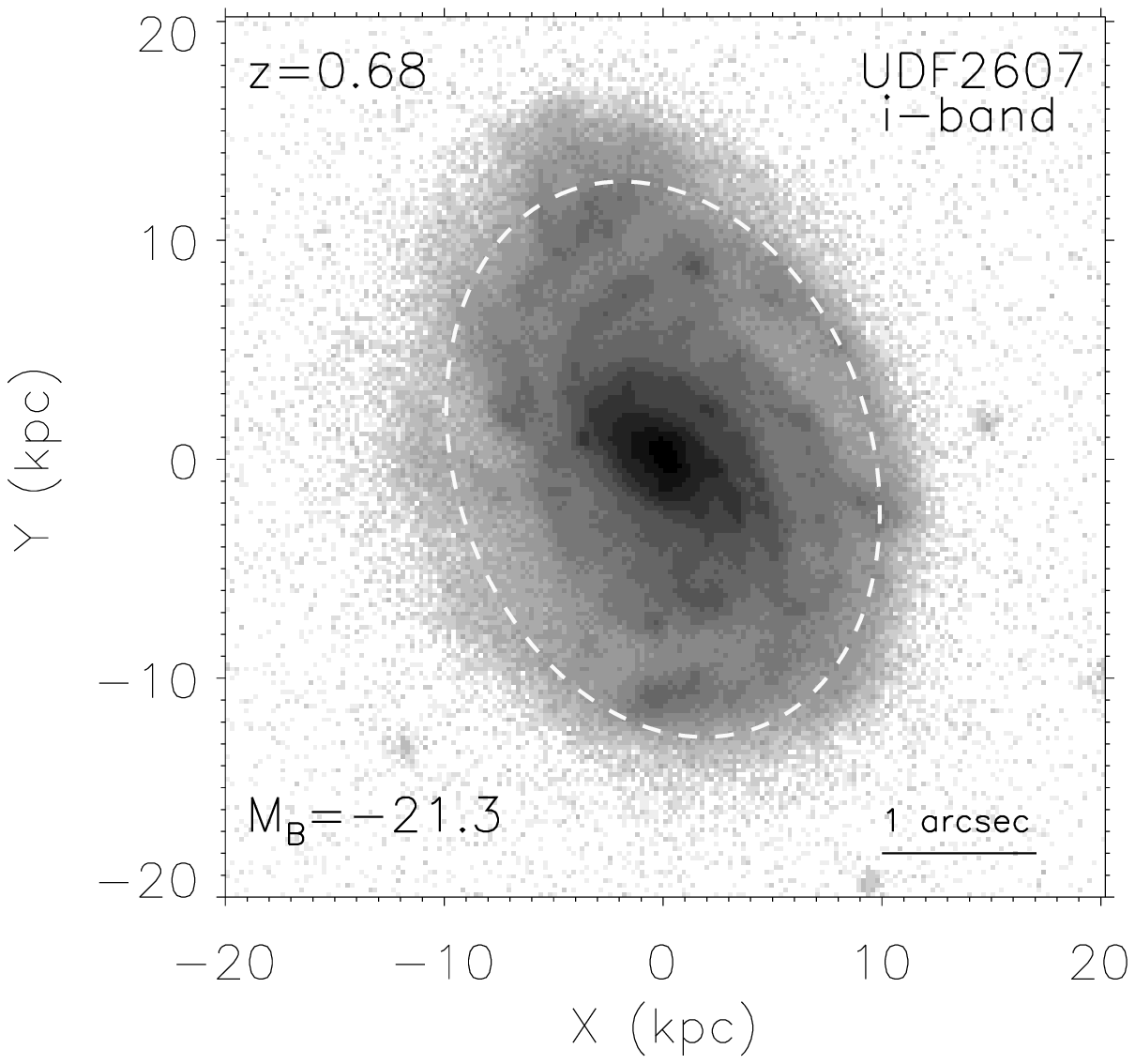}
\hspace{-3cm}
\vspace*{-0.7cm}
\includegraphics[width=10cm]{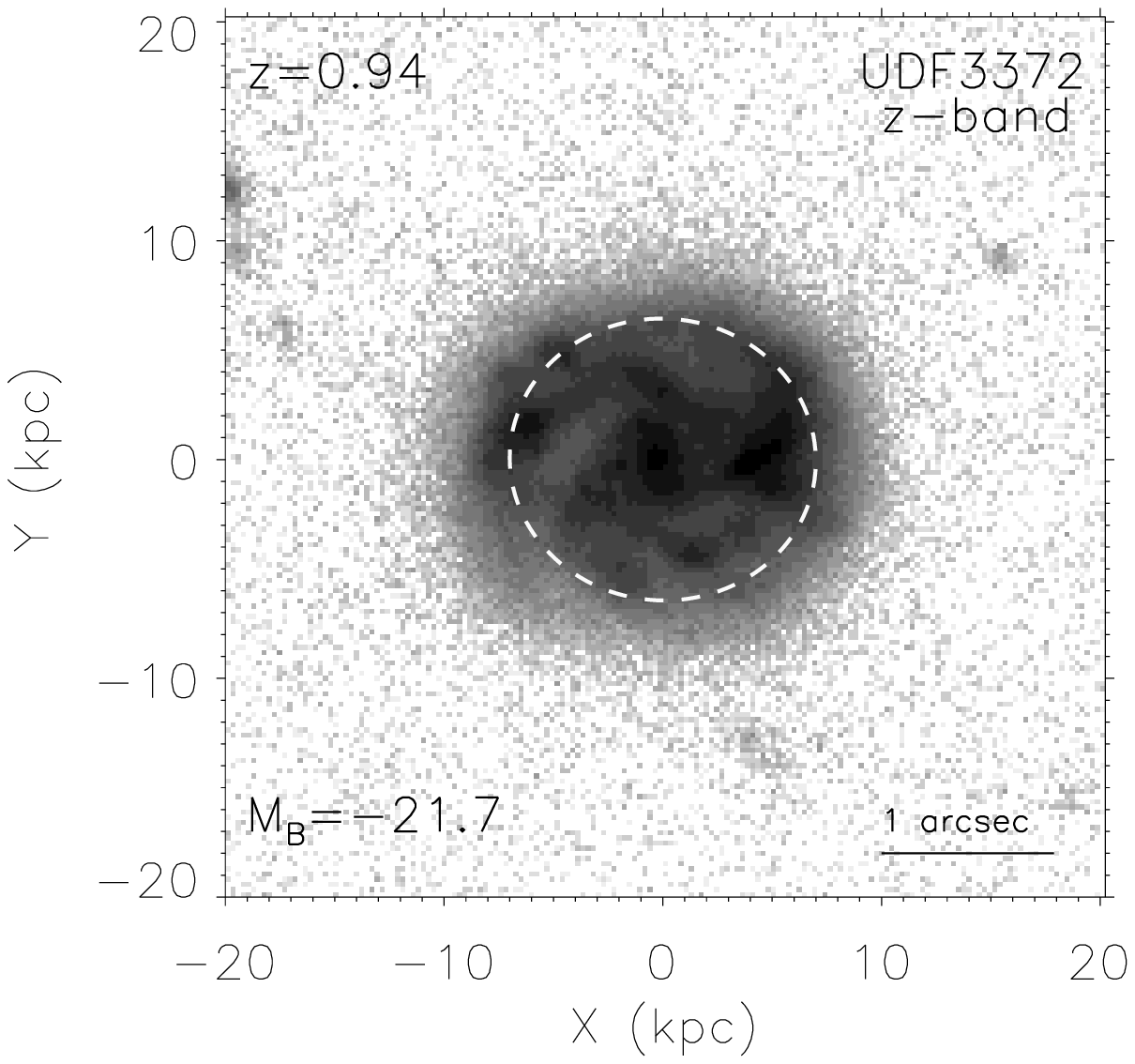}
\hspace*{0.1cm}
\includegraphics[width=7.5cm]{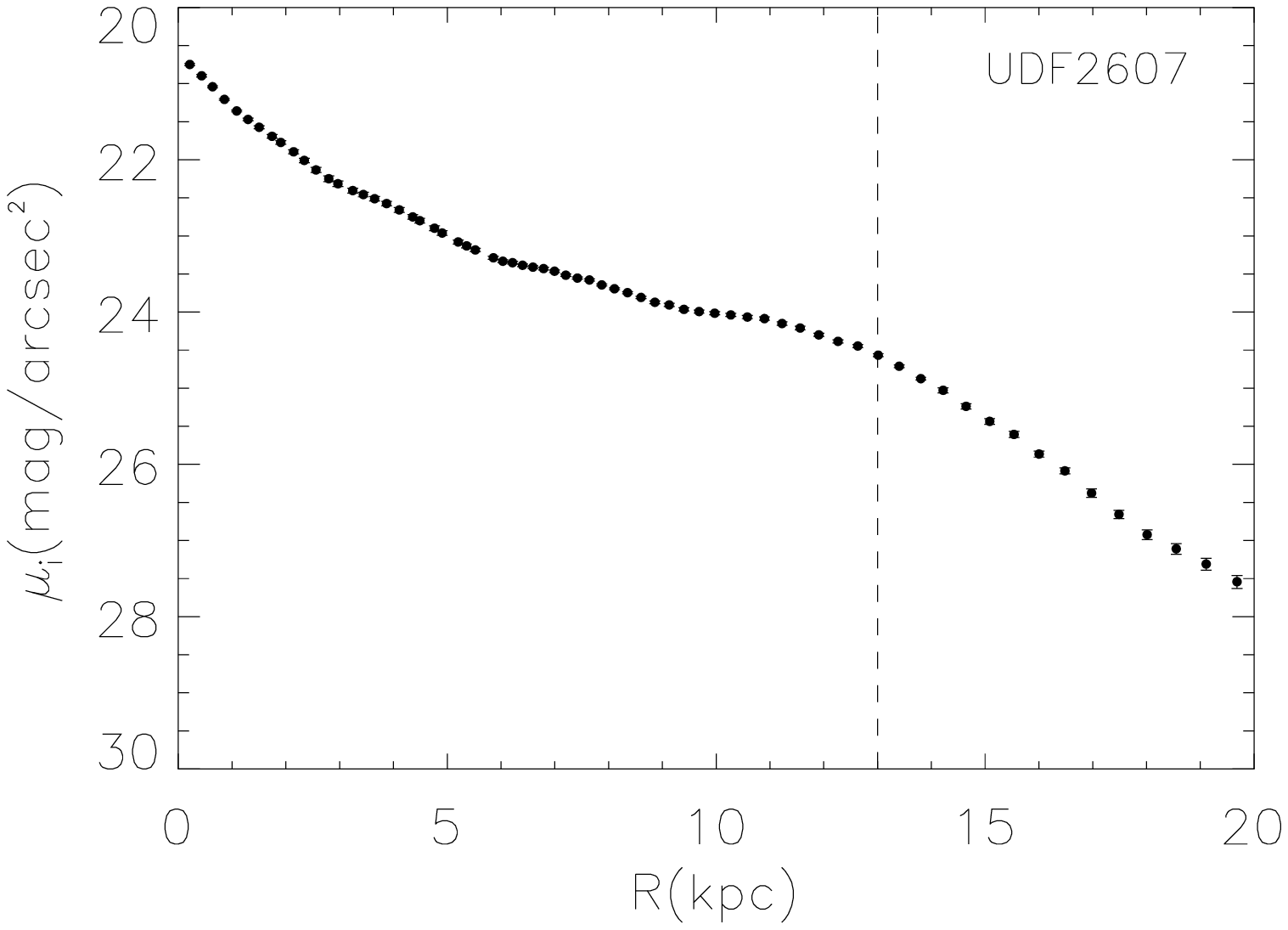}
\hspace{0.4cm}
\includegraphics[width=7.5cm]{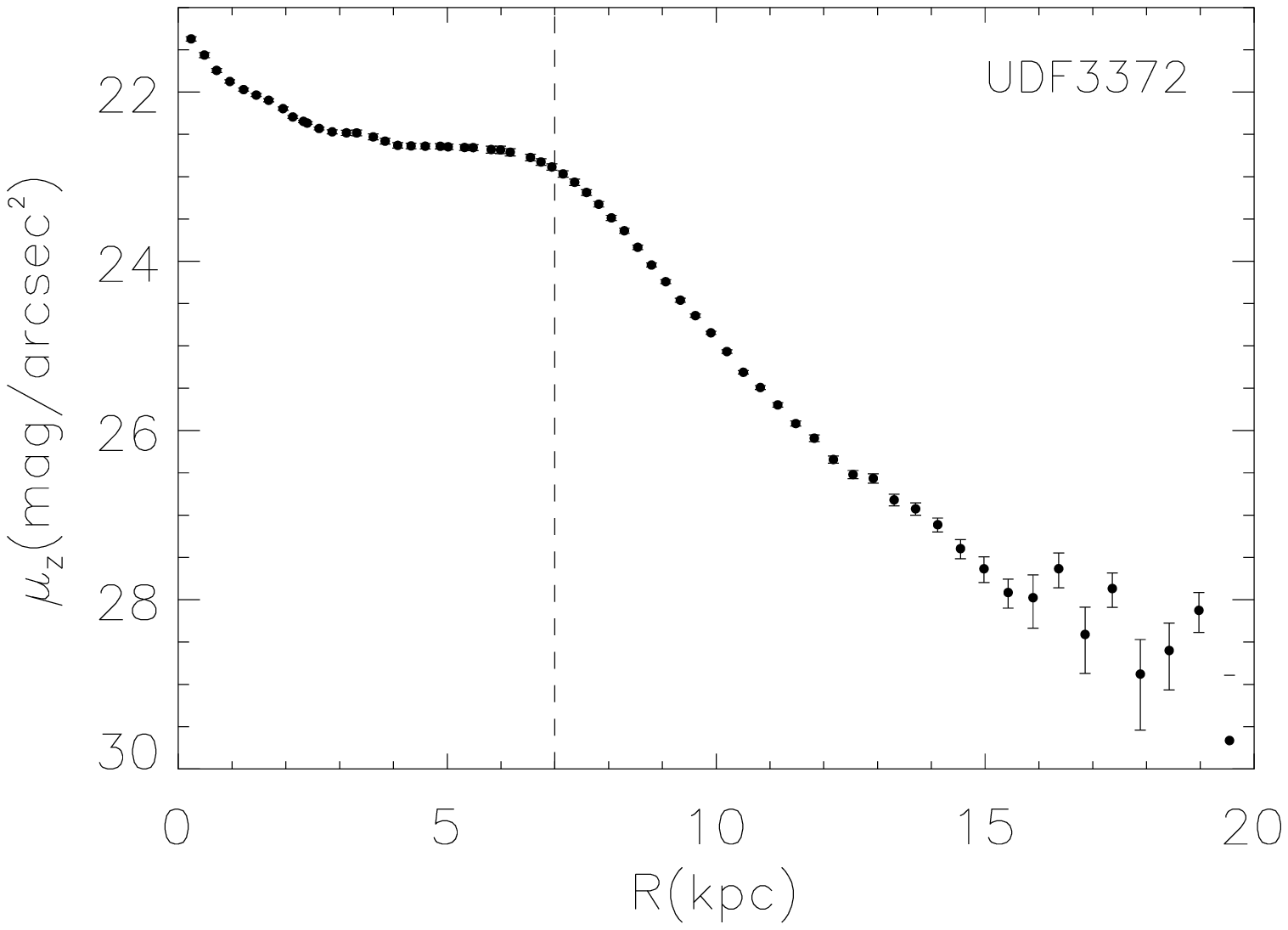}
\vspace*{-0.7cm}

\includegraphics[width=10cm]{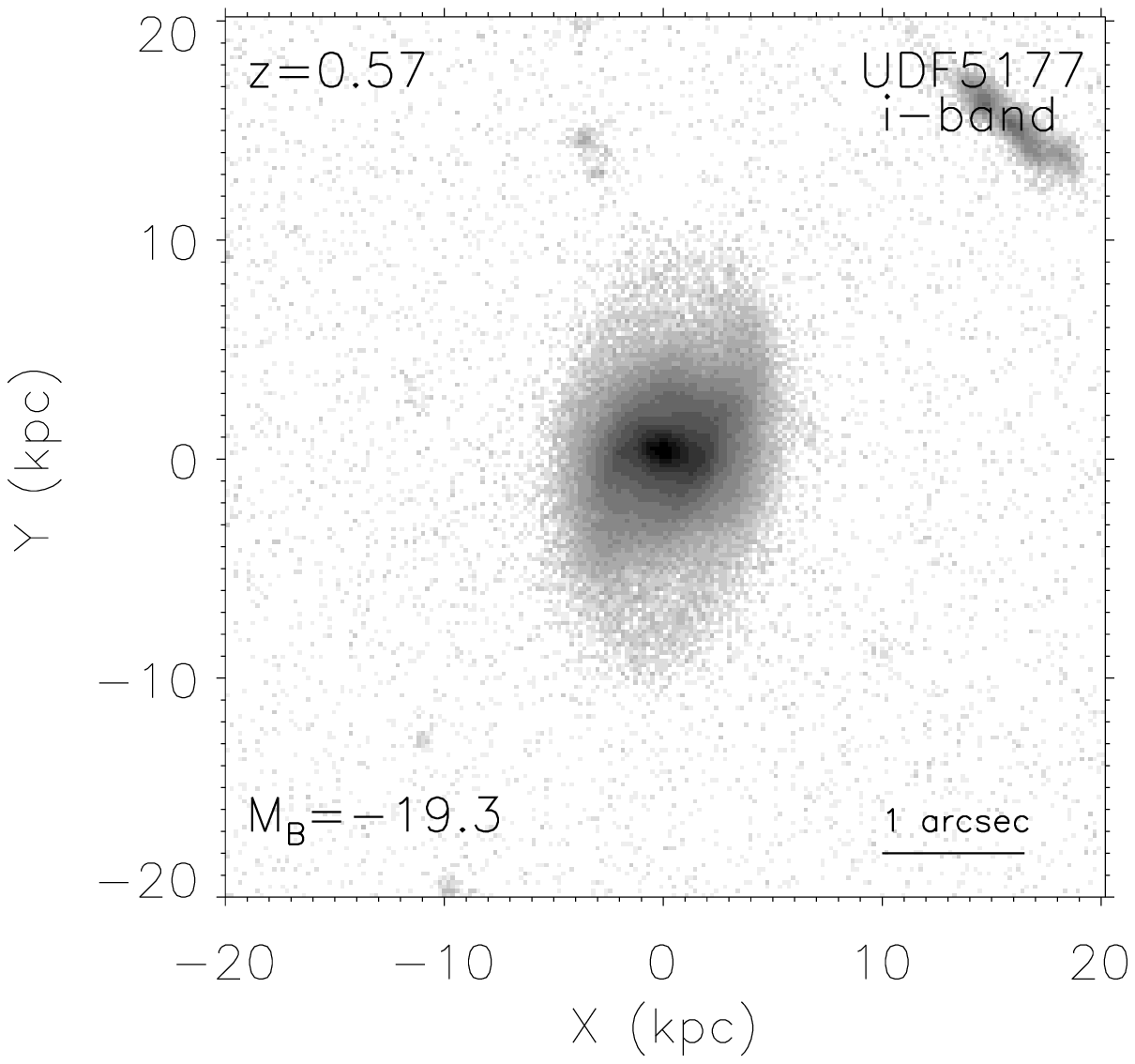}
\hspace{-3cm}
\vspace*{-0.7cm}
\includegraphics[width=10cm]{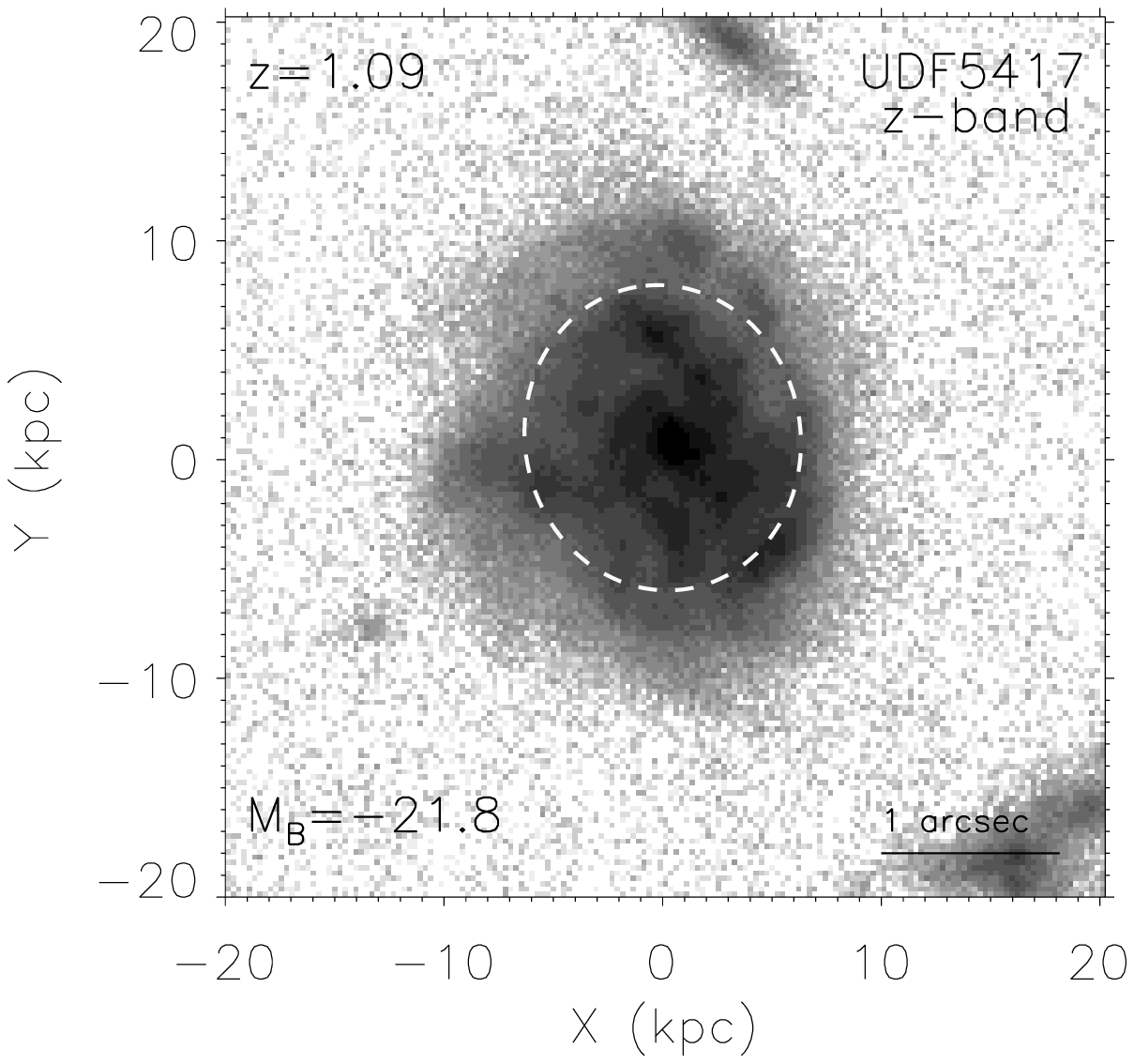}
\hspace*{0.1cm}
\includegraphics[width=7.5cm]{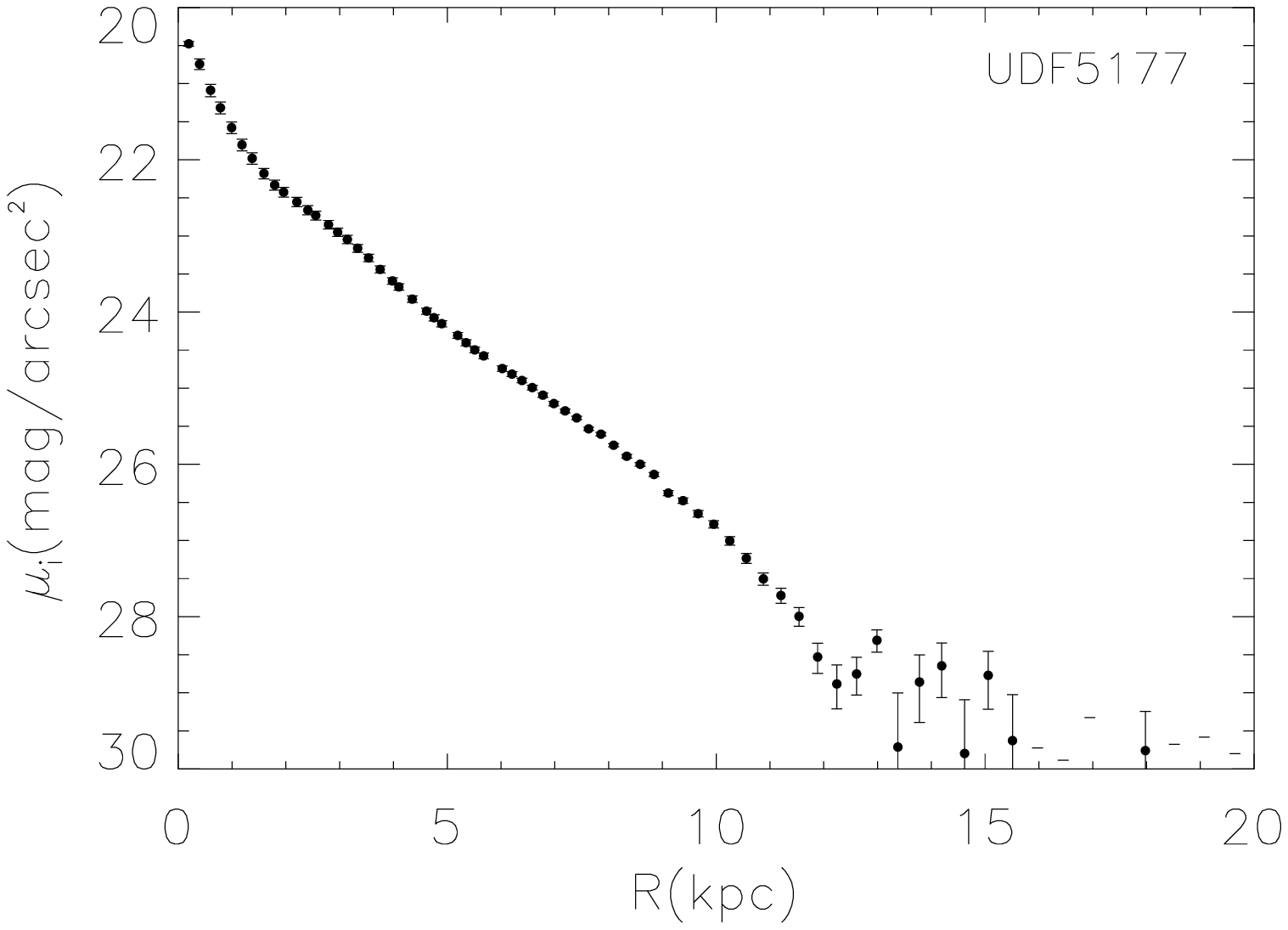}
\hspace{0.4cm}
\includegraphics[width=7.5cm]{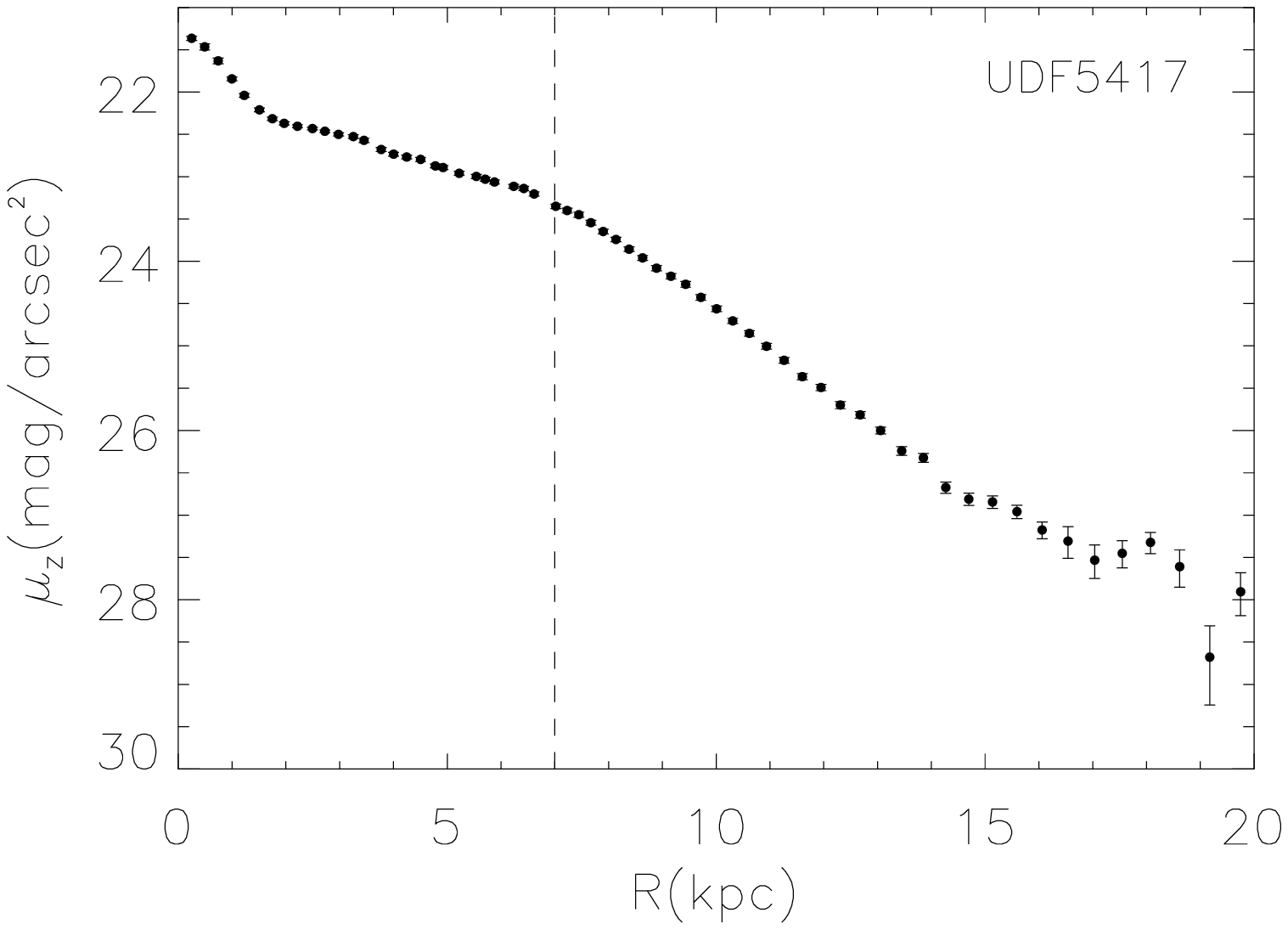}

\caption[]{\footnotesize{Surface brightness profiles and galaxy images  in the observed band
closest to the B--band restframe for different objects in our sample. The location of
the break radius has been overplotted on the images and on the profiles with a dashed
line. The surface brightness range shown in all the images expands from the peak of
the profile down to 27 mag/arcsec$^2$.}}

\label{sample} 
\end{figure}

The surface brightness profiles   were extracted  through ellipse fitting using
the ELLIPSE task within IRAF. Fitting ellipses over the whole galaxy produces
similar surface  brightness  profiles than those obtained through averaging
different image segments (Pohlen et al. 2002; P\'erez 2004). Initially the
ellipticity (E) and the position angle (PA) of the elliptical isophotes are
left as free parameters in order to determine the best set of E and PA
describing the outer disk. This is done for all the galaxies in the z--band.
The isophote fitting in the z--band  is less affected by additional structure
in the galaxies like spiral arms and prominent blobs of star formation and,
consequently, the above parameters are retrieved more accurately. We select the
E and PA at the radius where the mean flux of the best fitted free ellipse
reaches 1$\sigma$ of the background noise (this means $\sim$26.3 mag/arcsec$^2$
in the z--band in the AB system). We have  checked visually that for all the
cases this criteria is selecting an isophote well outside the inner region of
the disk. Once E and PA are determined we run again ELLIPSE with these
parameters fixed. In order to avoid contamination in our isophote fitting we
masked all the surrounding neighbors. Examples of the final surface brightness
profiles obtained are shown in Fig. \ref{sample}. 

From our sample of 36 galaxies there were 21 galaxies which show a truncation in the
surface brightness profiles. We have named these galaxies  Type DB (i.e.
DownBending). The exact details of how the position of the breaks were estimated are
explained in Pohlen \& Trujillo (2005), consequently here we explain only very
briefly the technique.  According to the observations, around the break the surface
brightness profile is well described by two exponential functions with  scalelenghts
h$_1$ and h$_2$, so we use the derivative of the surface brightness profile to
estimate the position of the transition. The position of the break is measured at the
radial position were the derivative  profile crosses the horizontal line defined by
(h$_1$+h$_2$)/2. The values obtained that way match very well with our estimations
done by eye. The  position of the break depends on the shape of the transition region
between both exponentials. A conservative estimate of the uncertainty of the position
 is $\sim$8\%. In addition, we do not find any systematic difference in the
position of the break at using the z--band or the band closest to the B--band
restframe.

There were 6 galaxies where we do not observe any signature of a break along the
profile. These galaxies are named Type I (Freeman 1970). We have another 9 galaxies
where the outer profiles are distinctly shallower in slope that the main disk
profile. We classify these as Type III following the notation by Erwin, Beckman \&
Pohlen (2005). We plan to study these objects in more detail in future papers.

\section{Discussion}

\begin{figure}
\plotone{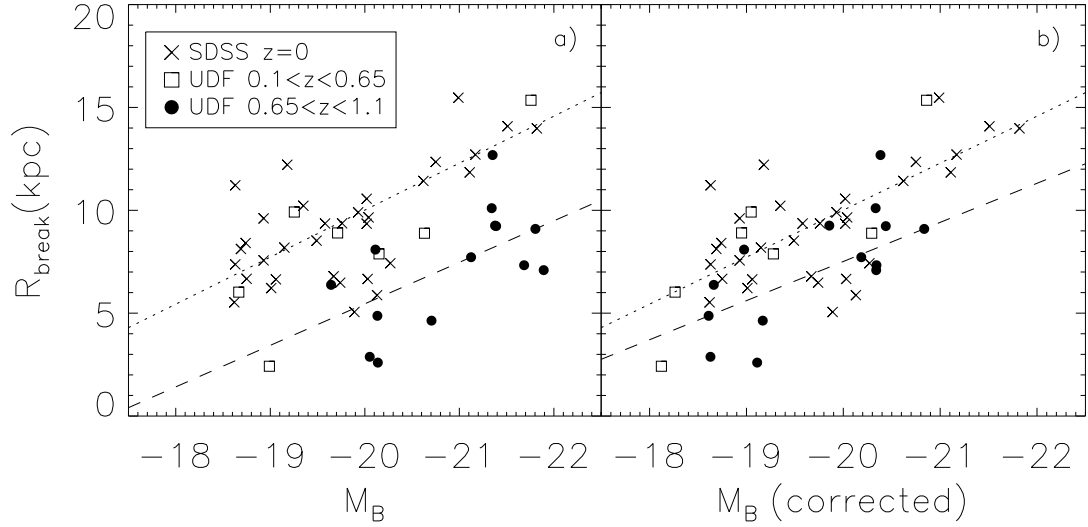}

\caption[]{Truncation radius versus absolute magnitude in the restframe B-band.
Panel a) shows the observed relation. Panel b) shows the relation after
correcting the absolute magnitude of the high--z objects accounting for the
observed mean surface brightness evolution of spiral galaxies since z$\sim$1
(Barden et al. 2005). After the surface brightness correction, the position of
R$_{break}$ in high--z galaxies seems to be systematically 1--3 kpc smaller
than in the local sample. The dotted lines corresponds to a linear fit to the
local relation whereas the dashed lines result from a fit to the galaxies with
z$>$0.65.}

\label{comparison} 
\end{figure}

In Fig. \ref{comparison}a we show the absolute B--band restframe magnitude
versus the position of the break for our high--z galaxy sample and, for
comparison, a volume selected local sample (Pohlen \& Trujillo 2005). The local
sample comprises  the 85 Sb--Sdm galaxies  from the LEDA catalogue having a
mean heliocentric radial velocity relative to the Local Group (corrected for
virgocentric inflow) $<$3250 km/s and M$_B$$<$-18.5 (AB system), with  useful
imaging data  available in the Sloan Second Data Release (DR2).   These
galaxies were selected to be face--on to intermediate inclined (e$<$0.5). This
sample represents the largest sample  ever used to homogeneously probe for
truncations using low inclination galaxies. From this sample we plot in Fig.
\ref{comparison}a the 35 galaxies which present a truncation. The truncation
radii of the local sample shown in Fig. \ref{comparison}a were estimated in the
SLOAN g--band.

The most simple description, a linear fit, of the local sample provides with
the following fit: R$_{break}$=-2.3$\times$M$_B$-35.7. We overplot this  in
Fig. \ref{comparison}a and Fig. \ref{comparison}b. On the other hand, galaxies
at z$>$0.65 present $\sim$4--5 kpc smaller truncation radii at a given
luminosity than the present--day galaxies (Fig. \ref{comparison}a). A linear
fit gives:  R$_{break}$=-2.0$\times$M$_B$-34.9. A concern is whether our
high--z galaxies are biased towards brighter surface brightness truncations
(and consequently towards smaller sizes). To test this we explored the surface
brightness distribution (in the observed band closest to the B--band
rest--frame) at the break positions. The distribution peaks at 24.1
mag/arcsec$^2$ with a scatter of 0.9 mag. Our faintest surface brightness at
the break is 25.5 mag/arcsec$^2$. We do not detect truncations between 25.5 and
27 mag/arcsec$^2$ (approximately our surface brightness detection limit). This
indicates our sample is not biased and we are probing all the truncations much
brighter than 27 mag/arcsec$^2$.

The stellar population of the galaxies are known to be younger at high--z and
consequently we expect that the galaxies were much brighter in the past than what
they are today. To check whether the evolution of the surface brightness alone is
able to explain the different distribution of the galaxies, we have applied a
correction on the  observed absolute magnitudes of the high--z galaxies using the
observed mean surface brightness evolution found in Barden et al. (2005) for
disk--like objects (i.e. we use d$<$$\mu_B$$>$/dz=-1.43$\pm$0.03). Under the
assumption of no evolution in the size of the objects this  corresponds  to the
same degree of evolution in the absolute magnitude (i.e.
M$_B$(0)=M$_B$(z)+1.43$\times$z).  The result of doing this is shown in Fig.
\ref{comparison}b. After the magnitude correction the position of the breaks for
the high--z population is still systematically smaller by 1--3 kpc at a given
``corrected'' luminosity. A linear fit results on
R$_{break}$=-1.9$\times$M$_B$-30.5. This  corresponds to an increase of $\sim$25\%
in the size of the inner disk since z$\sim$1. Although we are using the largest
samples in existence where the truncations have been explored systematically, the
estimation of the exact  evolution is limited by the small number statistic. It is
important, however, to note that this is the first time that the position of the
high--z stellar disk truncations is used to probe directly  the growth of  disk
galaxies with cosmic time.

Our results are consistent with a moderate inside--out formation of galactic
disks. This agrees with the conclusions obtained indirectly by Barden et
al. (2005) and Trujillo et al. (2004; 2005) analysing the evolution of the
luminosity and stellar mass--size relations since z$\sim$3.

\acknowledgments

We are very grateful to Marco Barden for kindly providing us with the GEMS
morphological analysis catalog. We acknowledge the COMBO--17 collaboration for
the public provision of an unique database upon which this study was based. We
thanks useful comments from Eric Bell, Peter Erwin and Renyer Peletier. The
referee comments helped to improve the paper. Based on observations made with
the NASA/ESA Hubble Space Telescope, which is operated by the Association of
Universities for Research in Astronomy, Inc, under NASA contract NAS5-26555.
This research was supported by a Marie Curie Intra--European Fellowship within
the 6th European Community Framework Programme.



\begin{deluxetable}{cccccc}
\tablecaption{Galaxy Sample}
\tablewidth{0pc}
\tablehead{\colhead{UDF ID} & \colhead{z$_{phot}$} & \colhead{M$_{B}$}
 & \colhead{Break Type} & \colhead{R$_{break}$} & \colhead{$\mu_{break}$} \\ 
 & & \colhead{(mag)} & & \colhead{(arcsec)} & \colhead{(mag/arcsec$^2$)}}

 \startdata

328  & 0.24 & -20.6 & DB  &   2.4 & 22.9 \\
900  & 0.45 & -20.8 & III &   --  & --   \\
901  & 1.00 & -20.1 & I   &   --  & --   \\
968  & 0.66 & -21.4 & DB  &   1.1 & 23.3 \\
1971 & 0.14 & -19.2 & DB  &   3.9 & 25.4 \\
2525 & 0.69 & -19.6 & DB  &   0.9 & 25.5 \\
2607 & 0.68 & -21.3 & DB  &   1.8 & 24.4 \\
3180 & 0.80 & -20.1 & DB  &   1.1 & 25.2 \\
3203 & 0.35 & -18.6 & I   &   --  & --   \\
3268 & 0.28 & -18.6 & DB  &   1.4 & 25.4 \\
3372 & 0.94 & -21.7 & DB  &   0.9 & 22.7 \\
3613 & 1.09 & -21.4 & I   &   --  & --   \\
3822 & 0.19 & -20.0 & III &   --  & --   \\
4142 & 0.67 & -20.6 & III &   --  & --   \\
4394 & 0.66 & -21.1 & DB  &   1.1 & 23.6 \\
4438 & 1.06 & -21.3 & DB  &   1.1 & 24.2 \\
4491 & 1.07 & -20.7 & DB  &   0.6 & 24.1 \\
4929 & 0.45 & -20.3 & III &   --  & --   \\
5177 & 0.57 & -19.3 & I   &   --  & --   \\
5268 & 0.61 & -19.0 & DB  &   0.4 & 23.3 \\
5417 & 1.09 & -21.8 & DB  &   0.9 & 23.4 \\
6821 & 1.07 & -20.1 & DB  &   0.5 & 23.9 \\
6853 & 0.79 & -19.2 & III &   --  & --   \\
6862 & 0.68 & -20.3 & III &   --  & --   \\
6974 & 0.61 & -20.1 & DB  &   1.2 & 25.1 \\
7112 & 1.00 & -20.0 & DB  &   0.4 & 23.2 \\
7556 & 0.63 & -21.7 & DB  &   2.3 & 24.4 \\
7559 & 0.93 & -20.4 & III &   --  & --   \\
8040 & 0.22 & -18.9 & I   &   --  & --   \\
8049 & 0.46 & -20.4 & III &   --  & --   \\
8125 & 1.07 & -20.5 & I   &   --  & --   \\
8257 & 0.57 & -20.1 & III &   --  & --   \\
8275 & 0.71 & -21.3 & DB  &   1.4 & 24.4 \\
8810 & 0.72 & -20.1 & DB  &   0.4 & 22.7 \\
9253 & 0.68 & -21.8 & DB  &   1.3 & 23.3 \\
9455 & 0.53 & -19.7 & DB  &   1.4 & 25.1 \\


\enddata

\tablecomments{Col. (1): Galaxy ID number according to the UDF catalog Col.
(2): Photometric redshift from COMBO-17 Col. (3): Absolute B--band restframe
magnitude from COMBO-17 (AB system) Col. (4): Surface brightness break type;
Type I: no break detected, Type DB: DownBending break, Type III:
anti--truncation (upbending). Col (5): Truncation Radius Col. (6) Observed
surface brightness at the break position in the  band closest to the
rest--frame B--band.} \label{table1} \end{deluxetable}


\end{document}